\documentclass[aps,showpacs,prl,twocolumn]{revtex4}

\usepackage{amsfonts}
\usepackage[T1]{fontenc}
\usepackage{graphicx}
\begin{document}
\title{Roughness suppression via rapid current modulation on an atom
chip}

\author{J.-B. Trebbia, C. L. Garrido Alzar, R. Cornelussen, C.I.
Westbrook, I. Bouchoule}
\address{Laboratoire Charles Fabry de l'Institut d'Optique, CNRS,
Univ Paris-Sud, Campus Polytechnique, RD128, 91127 Palaiseau cedex,
France}
\begin{abstract}
We present a method to suppress the roughness of the potential of a
wire-based, magnetic atom guide: modulating the wire current at a
few tens of kHz, the potential roughness, which is proportional to
the wire current, averages to zero. Using ultra-cold $^{87}{\rm Rb}$
clouds, we show experimentally that modulation reduces the roughness
by at least of a factor five without measurable heating or atom
loss. This roughness suppression results in a dramatic reduction of
the damping of center of mass oscillations.
\end{abstract}

\pacs{39.25.+k, 03.75.Be}

\maketitle

Atom chips, devices which trap and guide atoms with micro-fabricated
structures on a substrate, hold enormous promise for the
applications of cold atoms and for the exploration of new physical
regimes of degenerate gases ~\cite{FolmanRevue,ChipSpecialIssue}.
For applications, their small size facilitates the design of a large
variety of structures and
functionalities~\cite{ConveyorBelt,gunther:170405,WangMichelson,SchummDWell,treutlein:203005}.
For fundamental studies, the strong confinement possible on an atom
chip permits, e.g. the study of low dimensional gases. In particular
nearly one dimensional (1D) gases have been studied in the weakly
interacting regime on atom chips
\cite{EsteveFluctuations,trebbia:250403}, and the attainment of the
strongly interacting or Girardeau regime
\cite{Girardeau,tolra:190401,Parades,PhysRevLett.91.250402,Weiss1D}
on a chip is attracting significant experimental effort
\cite{reichel-2004-116}.

To exploit the full potential of these devices, atoms must often be
close to the material structures they contain.  This proximity
however, renders atom chips highly sensitive to
defects~\cite{Whitlock,sinclair:031603} which produce roughness in
the trapping potential. In the case of current carrying wires, most
of these defects are due to the wires
themselves~\cite{kraft,Hinds,leanhardt:100404,EstevePRArug,lukin}.
Advances in fabrication procedures have steadily improved wire
quality~\cite{sculturing,EsteveDWell}, but the roughness of the
trapping potential remains a problem. For example, it is a serious
hindrance for the attainment of the Girardeau regime, where the
atoms need to be tightly confined in the transverse direction and
weakly confined in the longitudinal direction.

In this paper we demonstrate a method, first suggested
in~\cite{kraft}, which nearly eliminates the longitudinal potential
roughness of a magnetic wire trap by rapid current modulation. This
idea is reminiscent of the TOP trap ~\cite{PhysRevLett.74.3352} in
the sense that the atoms see a time averaged potential. The wire
configuration here is such that the rough potential component
rapidly oscillates and averages to zero. We compare the potential
roughness for unmodulated (DC) and modulated (AC) guides (with the
same current amplitudes) and demonstrate a reduction factor of at
least 5 without noticeable atom loss or heating. This roughness
suppression results in a significant reduction of the damping of the
center of mass oscillations.

A simple atomic guide can be made by a current carrying wire and a
homogeneous field $B_{\rm bias}$ perpendicular to the
wire~\cite{FolmanRevue}: atoms are confined transversely and guided
parallel to the wire. Atoms, of magnetic moment $\mu$ in the low
field seeking state, feel a potential proportional to the absolute
value of the magnetic field. The guide is centered on a line where
the transverse field vanishes and the longitudinal potential $V(z)$
is given by $\mu |B_z^0+\delta B_{\rm z}^{\rm w}(z)|$, where $B_z^0$
is an external homogeneous field, typically of one Gauss, and
$\delta B_{\rm z}^{\rm w}(z)$ is the small, spatially fluctuating
field (few mG) created by current flow deformations inside the
micro-wires. Since $|B_{\rm z}^0|> \delta B_{\rm z}^{\rm w}(z)$ and
$\delta B_{\rm z}^{\rm w}(z)$ is proportional to the micro-wire
current, current modulation around zero results in a longitudinal
time-averaged potential in which the effect of $\delta B_{\rm
z}^{\rm w}(z)$ disappears.
 The instantaneous position of the guide is determined by
both the wire current and the transverse bias field $B_{\rm bias}$.
If only the wire current is modulated, the atom experiences strong
forces that eliminate the transverse confinement. One must therefore
also modulate $B_{\rm bias}$. In our experiment, this latter field
is also produced by on-chip micro wires whose low inductance permits
high modulation frequencies.

\begin{figure}
\scalebox{1}{\includegraphics{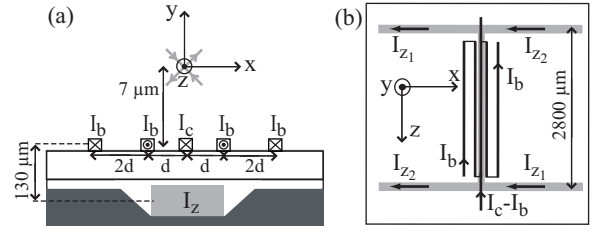}} \caption{Side (a) and top (b) views
of the atom-chip. The first wafer (dark grey) has 15~$\mu$m thick wires (grey). On
the second wafer (white), a five wire structure creates the atomic guide whose
roughness is studied ($d=2.5~\mu$m). Lengths not to scale.} \label{figure_schema}
\end{figure}

We use a two layer atom chip as sketched in Fig.\ref{figure_schema}. The lower
silicon wafer has a H structure of $15~\mu$m thick wires which carry the currents
$I_{\rm z1}$ and $I_{\rm z2}$. Atoms are initially loaded in an initial magnetic
trap (Z-trap) \cite{FolmanRevue}, which is created by setting $I_{\rm z1}= 3$~A,
$I_{\rm z2}=0$~A, and applying a transverse field (in the $x-y$ plane). After
evaporative cooling, we have a few times $10^4$ trapped $^{87}$Rb atoms in the
$|F=2,m_{\rm F}=2\rangle$ state at a temperature of the order of $1~\mu$K. Next, we
transfer the atomic cloud into a second trap (five-wire trap), where we perform the
potential roughness measurements for both DC and AC cases. This five-wire trap is a
combination of a transverse magnetic guide and a weak longitudinal confinement. The
guide is realized with five parallel micro-wires ($700\times 700~$nm$^2$ and $2$~mm
long in the $z$ direction) deposited on the upper silicon wafer. The wire geometry
is described in~\cite{EsteveDWell}, and shown in Fig.~\ref{figure_schema}. By
applying the currents $I_{\rm b} = \pm 15$~mA in the two outer pairs of wires and
$I_{\rm c} = \pm 13$~mA in the central wire, the resulting magnetic field vanishes
on a line (parallel to the $z$ axis), $7~\mu$m above the central wire. To prevent
Majorana losses, we add a constant and homogenous longitudinal field $B_{\rm
z}^0=1.8$~G. The weak longitudinal confinement is provided by the H-structure on the
lower chip with $I_{\rm z1}=I_{\rm z2}=0.4~$A (no current flows in the central wire
of the H).

\begin{figure}
\scalebox{1}{\includegraphics{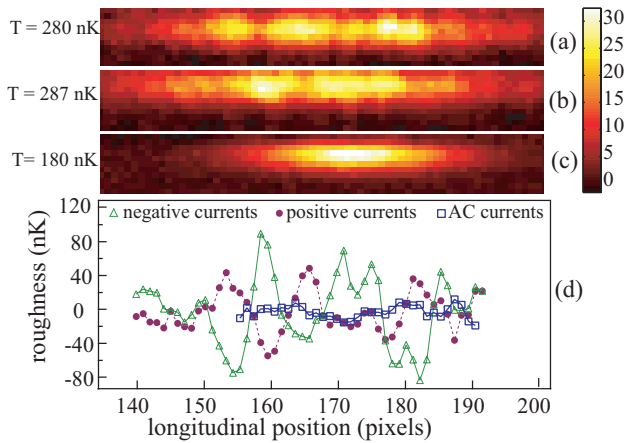}} \caption{(Color on line).
Absorption images of a thermal cloud for negative DC (a), positive DC (b), and AC
currents (c), respectively. Plotted is the number of atoms per pixel (pixel size is
$6 \times 6~\mu$m$^2$). In (d), the potential roughnesses are extracted from
longitudinal profiles using Maxwell-Boltzmann's distribution. }
\label{figure_roughness}
\end{figure}

First, we measured the roughness of the trapping potential for positive and negative
DC currents. After the transfer to the five-wire trap, we wait 600~ms to let the
sample thermalize in the presence of an evaporative knife  leading to about $5
\times 10^3$ atoms at around $280$~nK. To ensure that the two DC traps have the same
transverse position to within 500~nm, we compensate the transverse magnetic fields
to within $0.1$~Gauss. Figures~\ref{figure_roughness}(a) and
~\ref{figure_roughness}(b) show absorption images, averaged over 50 runs, of the
trapped cloud at thermal equilibrium for the two DC traps. In both situations the
cloud becomes fragmented, revealing the presence of potential roughness. Moreover,
the density maxima and minima are inverted when the currents are reversed, as
previously observed in ~\cite{kraft}. The time of flight (tof) for imaging is
$1.5$~ms. This delay allows the atoms to leave the vicinity of the five-wire
structure where diffraction of the imaging beam produces systematic errors in the
atomic density profile. The $1.5$~ms tof of the atoms partially washes out the
density modulation. This effect and the optical resolution contribute to a
resolution function with a rms width of $8~\mu$m modeled by a gaussian function.
Convolution with this gaussian gives a contrast reduction of 40\% for a wavelength
of $60~\mu$m, the typical spatial scale observed in Fig.~\ref{figure_roughness}. The
10 \% modulation in the figure is thus an underestimate of the actual potential
roughness. This smearing does not affect the relative measurements we present below,
except insofar as to reduce our signal to noise ratio.

To extract the potential $V(z)$ along the guide center from the
linear atomic density $n(z)$, we use the Maxwell-Boltzmann
distribution $V(z)=-k_{\rm B}T\ln[n(z)/n_0]$. This formula is valid
because the transverse oscillation frequency is independent of $z$
(to within 1\% over the cloud extent) and provided that the phase
space density is much smaller than unity. We deduce the temperature
from the transverse velocity distribution measured by the standard
tof technique.

The longitudinal confinement, produced by the H-shaped wire, is
expected to be harmonic within 6\% over the extent of the cloud. A
harmonic fit of $V(z)$ gives $\omega_{\rm z,DC}/2\pi = 7.1$~Hz.
Subtracting this potential from $V(z)$, we obtain the potential
roughness plotted in Fig.\ref{figure_roughness}(d) for both positive
and negative DC currents. The observed rms roughness amplitudes are
39~nK and 22~nK for negative and positive currents, respectively.
This asymmetry in the potential roughness may be due to residual
transverse magnetic fields which change the trap location, or to
residual noise in the imaging system.

We now turn to the study of the AC trap. Using phase-locked
oscillators at a frequency $\omega_{\rm m}/2\pi = 30$~kHz, we
produce the five-wire currents $I_{\rm c}(t) = \mathcal{I}_{\rm c}
\cos(\omega_{\rm m} t)$ and $I_{\rm b}(t) = \mathcal{I}_{\rm  b}
\cos( \omega_{\rm m}  t)$, where $\mathcal{I}_{\rm c}=13$~mA and
$\mathcal{I}_{\rm b}=15$~mA are the same currents as those used in
the DC cases. The  field $B_{\rm z}^0$ and the currents $I_{\rm z1}$
and $I_{\rm z2}$ are not modulated. Loading a thermal cloud in the
AC trap, the fragmentation is no longer visible in the absorption
image Fig.~\ref{figure_roughness}(c).

To give a more quantitative analysis, we begin with a theoretical description of the
AC trap. Since $\omega_{\rm m}$ is much larger than all the characteristic
frequencies associated with the atomic motion, the transverse and the longitudinal
dynamics are described by the one cycle averaged potential $\langle V(r,t) \rangle$.
Because $B_{\rm z}^0$ is large compared to any other magnetic field, $\langle V(r,t)
\rangle$ can be approximated by
\begin{equation}
\langle V(r,t) \rangle=\mu_{\rm B} \langle | B_{\rm z}(\textbf{r},t)| \rangle +
\frac{\mu_{\rm B}}{2 B_{\rm z}^0} \langle |\textbf{B}_{\perp}^{\rm
w}(\textbf{r},t)+\textbf{B}_{\perp}(\textbf{r})|^2 \rangle\cdot
\label{equation_expression_potentiel}
\end{equation}
Here, $\mu_B$ is the Bohr magneton, $B_{\rm z}(\textbf{r},t)$ is the
longitudinal magnetic field, $\textbf{B}_{\perp}^{\rm
w}(\textbf{r},t)$ is the transverse magnetic field of the five
wires, and $\textbf{B}_{\perp}(\textbf{r})$ denotes the transverse
field created by the H wires plus any uncompensated, homogeneous,
time-independent, transverse field. Since $\delta\textbf{B}_z^{\rm
w}$ oscillates, its contribution to $\langle |B_z(\textbf{r},t)|
\rangle$ averages to zero.

The cross term $\langle \textbf{B}_{\perp}^{\rm
w}(\textbf{r},t)\cdot\textbf{B}_{\perp}(\textbf{r})\rangle$
vanishes, so that $\textbf{B}_{\perp}(\textbf{r})$ only produces a
position dependent energy shift $\mu_{\rm
  B}|\textbf{B}_{\perp}({\bf r})|^2/2B_{\rm z}^0$.
It has a negligible effect on the transverse confinement since the
confinement created by the five-wires is very strong. On the other
hand, $|{\bf B}_\perp(\textbf{r})|^2$ affects the weak longitudinal
confinement. The length scale on which
$\textbf{B}_{\perp}(\textbf{r})$ varies is much larger than the
cloud extent, because it is produced by elements located at least a
millimeter away from the atomic cloud. It thus only contributes to a
curvature without introducing additional roughness. The  transverse
magnetic field created by the H-shaped wire is ${\bf
B}_\perp(z)=B^{'}_{\rm H}z~{\bf u_{\rm y}}$, where ${\bf u}_{\rm y}$
is the unit vector pointing along $y$. This field increases the
longitudinal frequency according to $\omega_{\rm z,AC}^2=\omega_{\rm
z,DC}^2+\mu_B B_{\rm H}^{'2}/(m B_{\rm z}^0)$. On the other hand,
the transverse  frequency is reduced by $\sqrt{2}$ compared to the
DC case. This factor is confirmed by our transverse frequency
measurements which give $(1.2 \pm 0.2)$~kHz for the AC trap,
$(2.2\pm0.1)$~kHz and $(2.1\pm0.1)$~kHz for DC negative and positive
currents, respectively.

We extract the potential in the AC trap from the measured
longitudinal profile for a sample at $T = 180$~nK. From a harmonic
fit we obtain a longitudinal frequency of 11.3~Hz, slightly larger
than the one in the DC trap, as expected. In
Fig.\ref{figure_roughness}(d), we compare the  measured roughness of
the AC trap to those obtained in the DC traps. In the AC
configuration, the roughness rms amplitude is reduced by a factor 4
and 7 with respect to the positive and negative DC cases. The
residual apparent roughness in the AC case is consistent with the
noise in the imaging system and so the mean reduction factor of 5 is
a lower limit.

\begin{figure}
\scalebox{1}{\includegraphics{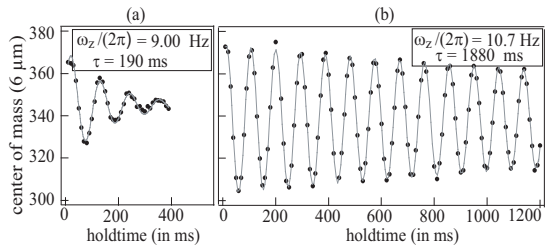}}
\caption{Center of mass position vs time for negative DC current (a)
and AC current (b) ($\omega_{\rm m}/2\pi= 30$~kHz). The solid line
is a fit to the function $Z(t)$ given in the text. The dramatic
increase in damping time in the AC trap illustrates the strong
reduction of the roughness.} \label{figure_damping}
\end{figure}

We also tested the reduction of the potential roughness by looking
at the evolution of the longitudinal center of mass oscillations
(CMO) as shown in Fig.\ref{figure_damping}. In a harmonic trap, the
CMO of an atomic cloud are undamped. In the presence of a potential
roughness, the oscillation period depends on the amplitude and the
dephasing between trajectories of different particles results in a
damping of the CMO. Figure~\ref{figure_damping} shows the dramatic
contrast between the AC and DC traps. We fit the oscillations to the
function $Z(t)=Z_0+Z_1e^{-t^2/\tau^2}\cos(\omega t +\phi)$, where
$Z_0$, $Z_1$, $\tau$, $\omega$ and $\phi$ are fitting parameters.
This function ensures the expected quadratic decrease of the
oscillation amplitude at small times for a damping due to trajectory
dephasing. The fitted damping time in the AC case, $\tau = (1.9\pm
0.1)$ s, is ten times larger than its value in the DC case. In fact,
the observed damping in the AC guide can be explained by the
anharmonicity of the H-shaped wire confinement alone. Thus, the data
are consistent with the absence of roughness.

To estimate an upper bound on the ratio of the AC to DC roughness
amplitudes, we performed one dimensional classical Monte Carlo
calculations for non-interacting atoms experiencing a rough
potential superimposed on a harmonic confinement. The calculations
are averaged over potential roughness realizations with the spectral
density expected for a single wire having white noise border
deformations~\cite{lukin,RealWorld}. We adjust the wire border noise
in order to recover the measured damping time in the DC trap (it
corresponds to an rms roughness of 80~nK). We find that the observed
increase of the damping time by a factor 10 requires a reduction of
the potential roughness by about $14$. A drawback of the model is
that it does not take into account atomic collisions, happening at a
rate on the order of the damping time. However, as collisions are
present in both the DC and AC configurations, we assume that the
above roughness reduction deduced from the calculation is still
relevant.

\begin{figure}
\scalebox{1}{\includegraphics{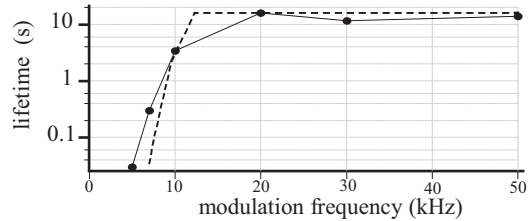}}
\caption{ Lifetime vs. $\omega_{\rm m}$ in the AC trap (dots). The
dashed line represents the result of Monte Carlo calculations with a
transverse magnetic field of $150$~mG for both x and y directions.}
\label{figure_lifetime}
\end{figure}

The modulation technique does not correct for all guide defects. By
considering only the time-averaged potential we neglect the
micro-motion of the atoms \cite{LandauMathieu} whose energy is
responsible for a residual roughness. This roughness is $10^{-8}$
times smaller than the one present in the DC guides for our
experimental parameters. In fact, other imperfections are more
important. First, the transverse component of the rough micro-wire
field produces meanders of the guide equipotentials (displacement of
about 2~nm) and a $z$-dependent modification of the transverse
trapping frequency (approximately 0.3\%). These two effects are not
averaged out in the AC guide. Second, the rough longitudinal field
$\delta\textbf{B}_z^{\rm w}$ appears in the expansion of $\langle
V(r,t) \rangle$ to orders that are neglected in
Eq.~(\ref{equation_expression_potentiel}). The resulting guide
imperfections are one order of magnitude smaller than those produced
by the transverse component of the rough field.

We also investigated  the dependence of the trap lifetime on
$\omega_{\rm m}$ as shown in Fig.\ref{figure_lifetime}. For
modulation frequencies above 15~kHz, the lifetime of the sample is
the same as in the DC case. For lower modulation frequencies, we
observe a decrease in lifetime that we attribute to instabilities of
the atomic trajectories in the modulated potential. For zero
unmodulated transverse magnetic field
$\textbf{B}_{\perp}(\textbf{r})$, the transverse atomic motion close
to the guide center is governed by the Mathieu
equation~\cite{LandauMathieu} and no instabilities are expected for
$\omega_{\rm m}>0.87\omega_{\perp,\rm DC}$, where
$0.87\omega_{\perp,\rm DC}/2\pi = 1.82$~kHz which is significantly
below $15$~kHz. We attribute this discrepancy to the presence of a
non vanishing transverse field.  Classical trajectory Monte Carlo
simulations, with fields of $150$~mG in both $x$ and $y$ directions,
reproduce well the measured lifetime as shown in
Fig.~\ref{figure_lifetime} (dashed line). This value is of the same
order as the accuracy of the residual transverse field compensation.
Within the explored range of frequencies (up to $50$~kHz), we have
not identified any upper bound for the modulation frequency. The
atomic spin should adiabatically follow the magnetic field
orientation to avoid spin flip losses. This condition is fulfilled
provided that $\omega_{\rm m}/2\pi \ll \mu_B B_{\rm z}^0/4\pi\hbar$,
where $\mu_B B_{\rm z}^0/4\pi\hbar\simeq 1.3$~MHz is the Larmor
frequency.

The temporal modulation of the rough potential can lead to a heating
\cite{bouchouleACtrap}. In our trap however, the expected heating
rate is very small. This is confirmed by our measured heating rate
of about $160$~nK/s which is close to the technical heating, also
observed in the DC trap. We have also loaded a Bose-Einstein
condensate in the AC trap without noticeable fragmentation.

The method presented here will enable new experimental explorations
on quantum gases using atom chips. We have already mentioned the
study of a 1D Bose gas in the Girardeau regime. Note that the
reduction of transverse trapping frequency by $\sqrt{2}$ can be
overcome by going closer to the wire. Second, the relative
insensitivity of the time-averaged potential to residual transverse
magnetic fields will enable us to revisit proposals of double well
potentials in static magnetic fields, such as
~\cite{Hinds,EsteveDWell}, which are very sensitive to residual
magnetic fields. Finally, this method may also allow the study of
atom dynamics in disordered
potentials~\cite{clement:170409,schulte:170411,lye:070401} since the
roughness strength can be tuned.

The authors thank A. Aspect, T. Schumm and J. Est\`{e}ve for
fruitful discussions. The Atom Optics group of Laboratoire Charles
Fabry is a member of the IFRAF Institute. This work was supported by
the EU under grants No.~ MRTN-CT-2003-505032 and No.~IP-CT-015714.

\bibliographystyle{prsty}

\end{document}